\documentclass[a4paper]{article}

\usepackage[T1]{fontenc} % Modern font encoding
\usepackage{float}       % For creating charts, graphs and schemes
\usepackage{helvet}      % Helvetica font for sans serif
\usepackage{mathptmx}    % Times font ("Word-like")
\usepackage{rsc}         % Loads natbib, etc.
\usepackage{setspace}    % For double-spacing
\AtBeginDocument{\doublespacing}

\newfloat{chart}{htbp}{loc}
\newfloat{graph}{htbp}{loh}
\newfloat{scheme}{htbp}{los}

\usepackage[version=3]{mhchem} % Formula subscripts using \ce{}

\usepackage{graphicx}
\usepackage{amssymb}
\usepackage{epstopdf}

\author{Scott T. Milner \\
Department of Chemical Engineering \\
The Pennsylvania State University \\
University Park, PA  16802
}

\title{Polymer crystal-melt interfaces\\
and nucleation in polyethylene}

\begin{document}

%%%%%%%%%%%%%%%%%%%%%%%%%%%%%%%%%%%%%%%%%%%%%%%%%%%%%%%%%%%%%%%%%%%%%
%% Make a title, of course
%%%%%%%%%%%%%%%%%%%%%%%%%%%%%%%%%%%%%%%%%%%%%%%%%%%%%%%%%%%%%%%%%%%%%
\maketitle

\begin{abstract}
Kinetic barriers cause polymers to crystallize incompletely,
into nanoscale lamellae interleaved with amorphous regions.
As a result, crystalline polymers are full of crystal-melt interfaces,
which dominate their physical properties.
The longstanding theoretical challenge to understand these interfaces
has new relevance, because of accumulating evidence
that polymer crystals often nucleate via a metastable, partially ordered ``rotator'' phase.
To test this idea requires a theory 
of the bulk and interfacial free energies of the critical nucleus.
We present a new approach to the crystal-melt interface,
which represents the amorphous region as a grafted brush of loops in a self-consistent pressure field.
We combine this theory with estimates of bulk free energy differences,
to calculate nucleation barriers and rates via rotator versus crystal nuclei for polyethylene.
We find rotator-phase nucleation is indeed favored 
throughout the temperature range where nucleation is observed.
Our methods can be extended to other polymers.  
\end{abstract}


\section*{Introduction}
\begin{figure}[htbp]
\begin{center}
\includegraphics[scale=0.65]{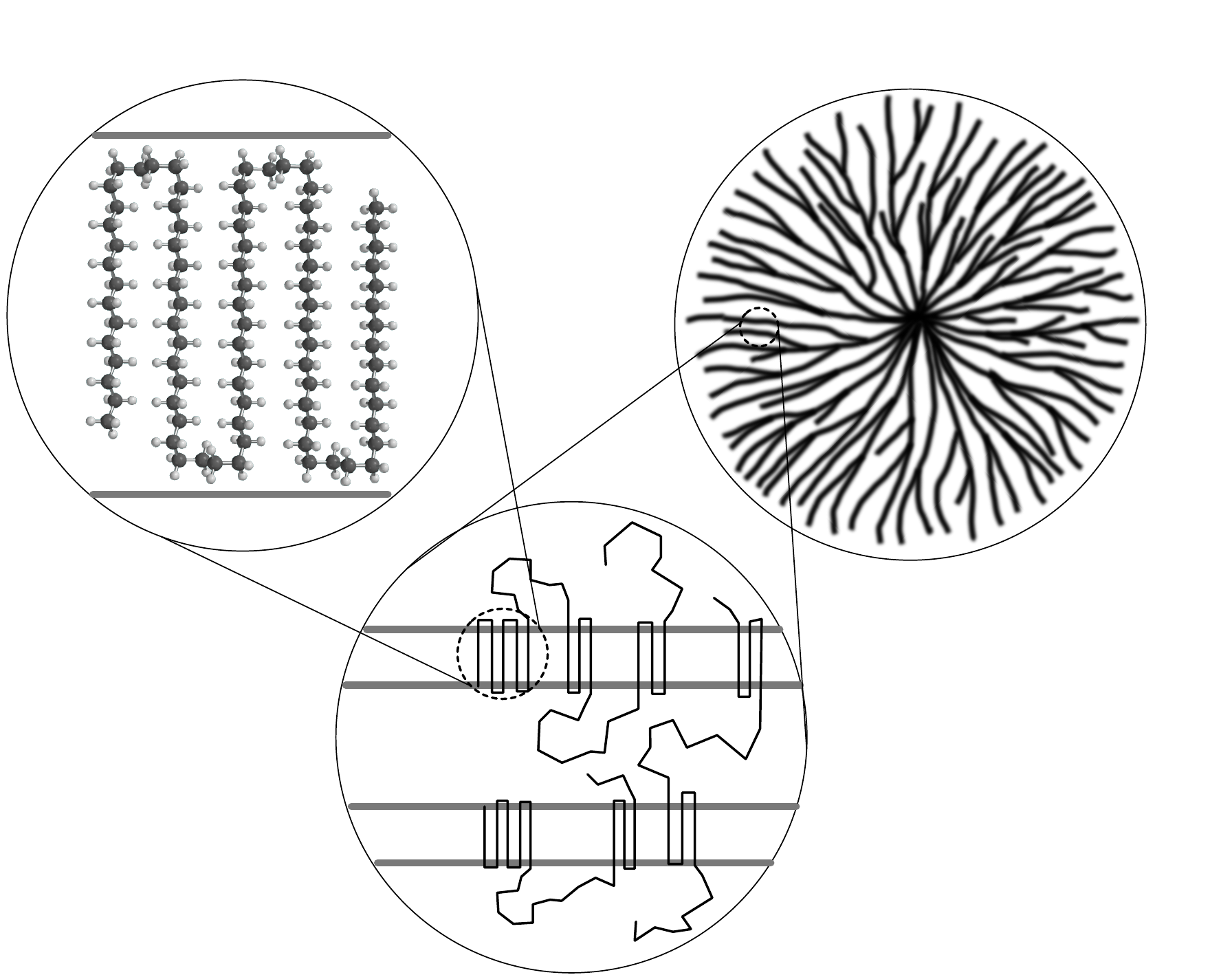}
\caption{``Multiscale'' schematic of semicrystalline polymers.  
Chain-folded crystalline lamellae of thickness $\sim$10nm 
lie adjacent to amorphous regions of comparable thickness; 
these are organized into micron-sized spherulites, 
consisting of lamellae emanating in all directions from a central nucleation site. }
\label{lam_spher}
\end{center}
\end{figure}

Polymer chain connectivity has an enormous influence 
on the structure of crystals and the kinetics of crystallization.
To begin with, crystallizable homopolymers are universally observed to be ``semicrystalline'' ---
that is, to consist of closely spaced lamellar crystallites of nanoscale thickness, 
separated by amorphous uncrystallized regions of comparable thickness  
\cite{WUNDERLICH:1976}.
(See Fig.\ \ref{lam_spher}.)

This crystalline lamellar structure is observed
even though the lowest free energy state of a crystallizable homopolymer
is certainly an extended-chain crystal, with no amorphous regions.
In such an crystal,  which can be grown only by heroic measures,
the long polymer molecules adopt a regular packing of linear conformations
(all-trans, or helices) in which the chains are all parallel.

The reason polymers routinely crystallize in a lamellar morphology is kinetic:
the time for chains in the random-coil configurations of a melt
to find their way to a fully extended state
is so long, that faster modes of crystal growth prevail
\cite{WUNDERLICH:1976}.
That faster mode is lateral growth of a crystal nucleus, 
by incorporating nearby chain segments as they are encountered by the growth front.
Only a portion of long chains will be ``combed out'' to parallel conformations 
as the lamella grows past.

The result of this lamellar growth is that on either face of the lamella,
all the snarls and tangles that were combed out from the crystallized region
accumulate in as yet uncrystallized material.
These accumulated entanglements, 
as well as the limited space between adjacent lamellae,
eventually inhibit further crystallization.

As a consequence, crystallizable polymers are full of crystal-melt interface.
This has dramatic consequences for the material properties.
The high degree of plastic deformation of solid polymers
depends on the existence of the amorphous regions
and their connectivity through chains to adjacent lamellae.

If the chains are so short that there are insufficient ``tie chains'' between lamellae,
the material becomes brittle
\cite{PHILLIPS:1976p5162,FISCHER:1984p5129}.
(For very short unentangled polymer molecules,
the above kinetic considerations, leading to lamellar crystals 
with amorphous material sandwiched between, no longer apply. 
Polyethylene of less than 50 carbons or so is alkane wax, and crystallizes completely;
equilibrium crystal-melt interfaces for such short chains consist mostly of chain ends,
and are not our focus in this paper.)

\subsection*{Previous theories}
The structure of polymeric crystal-melt interfaces has long been a challenge to theory.
A generation ago, there was spirited debate %changed
over how amorphous polymer chains emerging from a crystalline lamella
reentered the lamella at some other location.
The contending views were the ``adjacent reentry'' model, 
which held that emerging chains predominately formed tight loops
and re-entered the crystal right next to where they exited
\cite{HOFFMAN:1970};
and the ``switchboard'' model, 
which said that emerging chains made fluffier, more random loops
and typically re-entered the crystal at distant locations
\cite{YOON:1977p5203,FLORY:1978p5207}.
% dropped sentence about age of controversy

Attempts were made to compute conformational probabilities
for the amorphous chain loops at the crystal-melt interface,
but they suffered from various shortcomings.
Either chain connectivity was not properly respected
\cite{YOON:1984p4198,FLORY:1984p4154,MARQUSEE:1986p4160,KUMAR:1989p4166,MARQUSEE:1989p4162,KUMAR:1989p4167},
or the constraint of constant melt density was not enforced,
in so-called ``Gambler's Ruin'' models of noninteracting chains at interfaces %%added
\cite{DIMARZIO:1980p5126,GUTTMAN:1981p5137,GUTTMAN:1982p5210}.
%% added references to Muthu/Sommer, and simulations
Models of the Gambler's Ruin type have been used more recently
to study crystallization of single chains in solution
\cite{Muthukumar:2003p6492},
and the shapes of single crystals composed of many chains
\cite{Sommer:2006p6584}.
But these works share all the shortcomings of their antecedents ---
they assume noninteracting, fully flexible, Gaussian chains ---
and so can only qualitatively describe the melt-crystal interface.

More recently, Monte Carlo simulation techniques have been fruitfully applied
to study the melt-crystal interface
\cite{Balijepalli:2000p3338,Gautam:2000p3340,Hutter:2006p3285}.
These united-atom simulations give detailed information on chain conformations,
internal stresses, and potential energies of an equilibrium interface.
However, these simulations are not able to predict interfacial free energies, 
a key parameter in nucleation theory, discussed below.
%% end of addition

Considerable work has also been done %%changed
on the properties of grafted polymers at interfaces --- ``polymer brushes''
\cite{SCHEUTJENS:1979p5211,COSGROVE:1987p5212,MILNER:1990p37}.
In this paper, I will use the modern theory of polymer brushes
to provide a definitive theory of the crystal-melt interface.
With this theory, predictions can be made for the interfacial structure,
free energy, fraction of adjacent reentry, and other physical properties of interest.

\subsection*{Nucleation}
There is a new reason to be interested in the structure 
of the polymer melt-crystal interface, 
which again relates to an old debate,
as to the mechanism for nucleation of polymer crystals.

The classical theory of nucleation in an undercooled, metastable liquid phase
posits a small `nucleus' --- a tiny crystal, that may or may not grow --- 
which comes into existence by random fluctuations
\cite{TURNBULL:1961p5200}.
(See Fig.\ \ref{critical_nucleus}a.)
The driving force for the nucleus to grow is that its free energy 
is lower than that of the surrounding melt.
However, the interface between the nucleus and melt has an interfacial tension,
which inhibits growth of small crystals.
The free energy as a function of crystal size is the sum of the bulk and surface terms,
with a maximum, corresponding to the ``critical nucleus'', 
beyond which further increases in size actually decrease the total free energy. 

For a cylindrical nucleus, appropriate to a lamellar crystal, 
the nucleation barrier is
\cite{UHLMANN:1975p5201,OLIVER:1975p5202}
\begin{equation}
\Delta G^* = {8 \pi \gamma_s^2 \gamma_e \over (\Delta S \Delta T)^2 } %% corrected
\label{nucl_barrier}
\end{equation}
in which $\Delta S$ is the transition entropy, $\Delta T$ the undercooling, 
and $\gamma_e$ and $\gamma_s$ are respectively 
the interfacial tensions on the ``end'' and ``side'' of the cylindrical nucleus.

The question arises:  what phase nucleates?  
The naive answer is, the most stable phase; 
but Ostwald's ``rule of stages'' reminds us that in matters of kinetics,
victory belongs to the swift --- the phase with the lowest nucleation barrier nucleates first
\cite{OSTWALD:1897p289}.
After nucleating, this phase may grow, or convert to the most stable phase.
This phenomenon is commonly observed and of great practical importance in metallurgy.

Regarding nucleation in polymers, long ago the suggestion was made 
that the critical nucleus consisted not of a small polymer crystal,
but rather a collection of more-or-less parallel chain segments 
gathered together from the melt like a sheaf of wheat
\cite{FLORY:1962p5309}.
This hypothetical structure for the critical nucleus, known as a ``fringed micelle'',
was rebutted by the argument that the fringed micelle would not have a strong bulk driving term
(not being a crystal with its favorable free energy), 
while still having an interface made of stretched and distorted ``fringe'' chains
\cite{ZACHMANN:1969p5180,LAURITZEN:1978p5196,GUTTMAN:1980p5199}.

\subsection*{Rotator phases}
\begin{figure}[htbp]
\begin{center}
\includegraphics[scale=0.55]{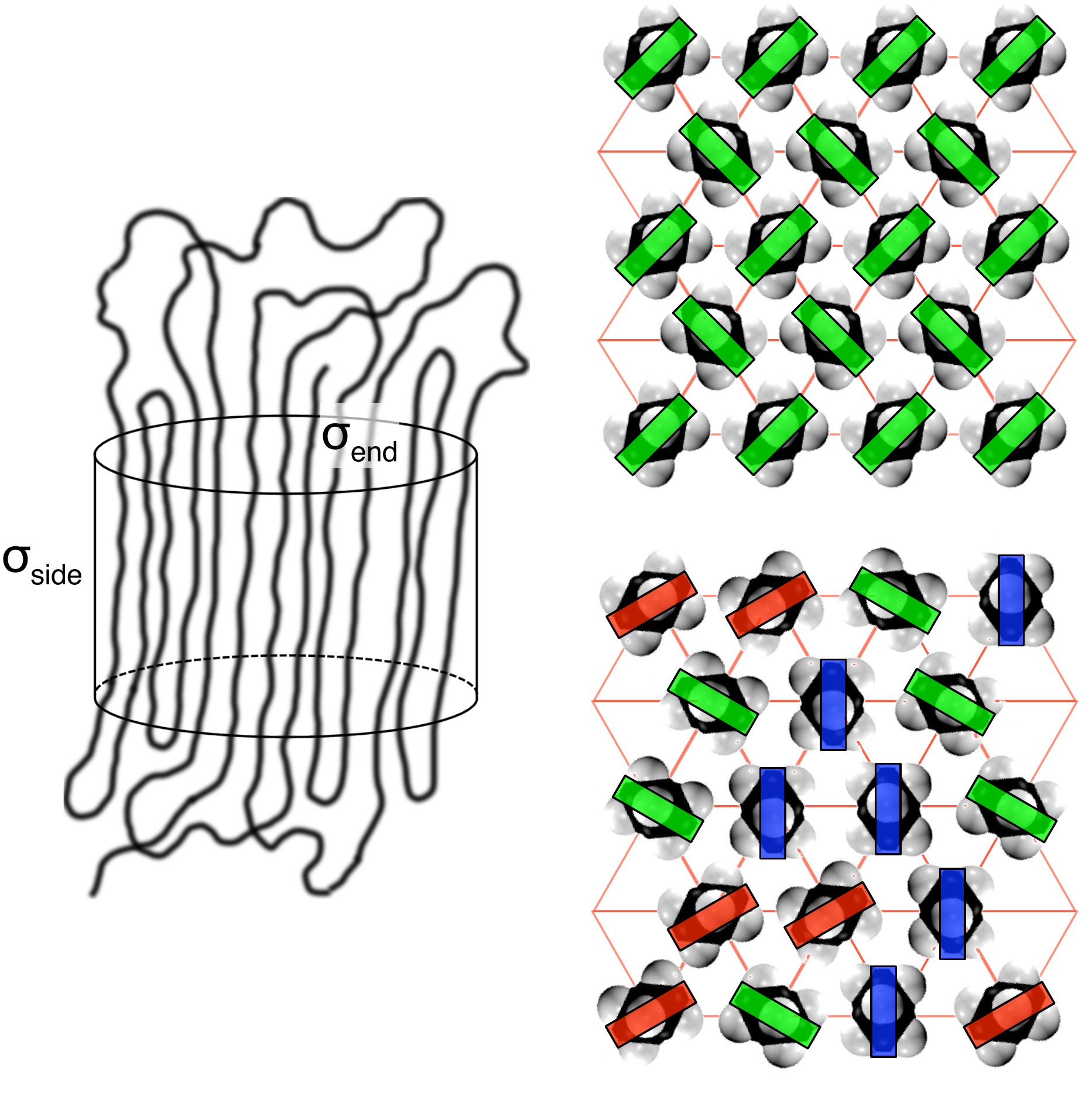}
\caption{a) Sketch of a nucleus for crystallization, consisting of a cylindrical bundle of parallel chain stems, bounded by a ``fringe'' of amorphous loops and chain segments into the surrounding melt.  b) Top view of the crystalline and rotator RII phases of normal alkanes (oligomeric PE).}
\label{critical_nucleus}
\end{center}
\end{figure}

However, strong evidence has recently accumulated 
that polyethylene nucleates via a metastable partially ordered phase, %% changed
called a ``rotator'' phase
\cite{RASTOGI:1991p7767,RASTOGI:1991p7765,Kraack:2000p2178,Kraack:2001p3329,Strobl:2000p2183}.
In a rotator phase, chains are parallel, 
and packed on a regular lattice in the plane normal to the chain direction,
but retain some freedom to rotate about their own axes.
They therefore have a somewhat lower areal density of chain stems
than does the crystal, which will be crucial for this paper.
(See Fig.\ \ref{critical_nucleus}b.)
Rotator phases are stable in linear alkanes with 25--60 carbons or so
(basically short pieces of polyethylene),
for a narrow range of temperatures between the crystal and melt phases
\cite{Sirota:1997p2176},
and are stable at elevated pressures and temperatures in polyethylene
\cite{Ross:1977p2185}.

The evidence for nucleation via rotator phases in polyethylene 
is circumstantial, but quite suggestive.
First, time-dependent X-ray scattering in linear alkanes shows
that a transient rotator phase appears during a quench into the crystal phase
\cite{Sirota:2000p2172,Sirota:1999p3188}.
%%odd-even effect added
Second, the quench depth required to achieve a given nucleation rate in linear alkanes
shows a smooth dependence on carbon number,
only if the quench is assumed to be with respect to the {\it rotator} phase melting temperature
\cite{Kraack:2000p2178}.

Also, in polyethylene with a bit of comonomer,
the comonomer sidechain is known to be rejected from the crystal phase;
only uninterrupted ethylene sequences crystallize. 
Indeed, the melting temperature of such crystals is found to be suppressed by the rejected comonomer, 
much as the osmotic pressure of salt suppresses the melting temperature of ice 
(from which the salt is rejected).

% melting line description added
This effect is evident in the ``melting line'' data of Strobl,
who presents melting temperature $T_m(d)$ versus lamellar thickness $d$
for a series of ethylene-octene copolymers.
The melting line, well described by the Gibbs-Thompson relation
\begin{equation}
T_m(d) = T_m(\infty) - 2 \sigma/d \Delta S,
\end{equation}
is shifted to lower temperatures with increasing octene content.

% crystallization line description changed
However:  the presence of a bit of comonomer has no effect on the ``crystallization line'',
which reports the lamellar thickness versus crystallization temperature $T_x$
\cite{Grasruck:2003p2171,AlHussein:2002p2182,Strobl:2000p2183}.
This suggests a picture in which lamellae nucleated into a mesophase,
then thickened (because chains in the mesophase are more mobile)
until converting at some larger thickness to the crystal phase.
No comonomer effect is observed,
because the comonomer was already expelled from the mesophase
\cite{Sirota:2007p8452}.

So the ``fringed micelle'' idea of polymer nucleation
may actually correspond to the current hypothesis that polyethylene,
and perhaps other polymers as well, nucleate via rotator phases.
In the context of Ostwald's Rule,
rotator phases would be favored to nucleate
if they had competitive bulk free energy to that of the crystal
(which is suggested by their being stable in a narrow range of temperatures for alkanes),
as well as a lower surface tension against the melt than that of the crystal.
So to predict what phase nucleates polyethylene,
we need a theory for the free energy difference between rotator and crystal phases,
and a theory for the interfacial free energy of both ordered phases against a melt.

\section*{New theory for melt-crystal interfaces}
Thus we again meet the question, what is the structure and free energy of the interface
between an ordered polymer phase (crystal or rotator) and the melt?
In the language of grafted polymer brushes,
the carpet of amorphous chain loops at the crystal-melt interface
is a ``pseudobrush'', first studied in the context of adsorbed polymers at interfaces
\cite{Aubouy:1996p2166} .
By pseudobrush, we mean a polydisperse brush of loops,
in which the length distribution of the loops is determined not by a chemist
but by the system itself, as it seeks to minimize its free energy subject to constraints.

For the crystal-melt interface pseudobrush, 
the constraints are 1) constant local ``melt'' density in the amorphous region, %% changed
and 2) an areal density of ``injected'' chains emerging from the crystal
set by the crystal structure (which we regard as unperturbed by the interface).
Subject to these constraints, the system determines the best loop-length distribution,
that maximizes configurational entropy and minimizes chain bending.
That is, the system must resolve the ``crowding paradox'':
the injected chains are at much too high an areal density
to make random walks in the amorphous region --- 
if they did so, they would overfill space.
So some of the chains must make tight loops 
(which cost bending energy) to reenter the lamella,
so that other chains may take more nearly random configurations
to gain entropy, and thus interpolate between the strongly oriented chains 
of the ordered phase, and an isotropic melt region well above.

To simplify the theorist's task, we make the following assumptions:
\begin{itemize}
\item Chains of infinite length (no free ends);
\item Semi-infinite melt region above the crystalline lamella (only one interface);
\item Interface in equilibrium at the melting point (can use equilibrium statistical mechanics).
\end{itemize}
These assumptions may seem at cross purposes with studying nucleation ---
infinite-length chains will take forever to equilibrate,
and how will the results be relevant to the formation and growth kinetics of a critical nucleus?
However, our equilibrium estimate will be a lower bound on the interfacial tension
of an incompletely equilibrated structure;
as well, we can argue that the residual effects of distant large loops
on the interfacial free energy will be minimal.
%% added regarding assumptions:
Free end effects are a small perturbation in a typical high molecular weight polymer sample;
tie chains are likewise a small perturbation, since estimates from GamblerÕs Ruin models indicate 
that only a few percent at most of stems give rise to tie chains.

\subsection*{Self-consistent field theory}
We use self-consistent field theory to describe melts of polymer chains 
in inhomogeneous environments, such as for chains attached to an interface
\cite{SCHEUTJENS:1979p5211,COSGROVE:1987p5212,MILNER:1990p37}.
In SCFT, we consider each chain to interact with others 
only through the hydrostatic pressure field,
which is adjusted (self-consistently) to enforce the constraint of constant melt density.
For a end-grafted polymer layer or ``brush'' at reasonably high coverage (chains per area),
the hydrostatic pressure is high near the grafting surface,
to drive the chains to stretch and thereby avoid overfilling space near the surface.

In cases where chains are ``strongly stretched'' well beyond their random-coil dimensions,
elegant analytical approximations exist for calculating chain conformations,
pressure profiles, and interfacial free energies
\cite{MILNER:1989p96,MILNER:1991p63,Aubouy:1996p2166}.
However, for the crystal-melt interface, we expect that many of the loops will be quite short,
and so the analytical strong-stretching approach would not be valid.

There are several equivalent formulations of SCFT
\cite{MATSEN:1994p5474},
but for present purposes the simplest is to consider chains 
on a semi-infinite simple cubic lattice, with the crystal face as boundary.
The lattice version of SCFT has the advantage 
of naturally imposing a short-distance ``monomer cutoff''.
(Bending energies can also be easily added, as we shall see below.)
Lattice SCFT is particularly tractable in problems such as a grafted layer, 
in which the potential only varies in one dimension.

Of course, we may consider pseudobrushes at different coverages.
One important limit corresponds to the pseudobrush we form
by cutting with a plane an isotropic melt of infinite-length chains.
Each half-space forms a pseudobrush of polydisperse loops,
in this case with constant hydrostatic pressure, 
so that the loops are isotropic random walks above an absorbing boundary.
The coverage for this pseudobrush on a simple cubic lattice is 1/6,
since for all the lattice sites adjacent to the boundary plane,
the probability for the chain passing through there to be headed across the plane is 1/6.
(This is true even for semiflexible chains, for which bends are penalized.)
The isotropic pseudobrush should be the minimum free energy state,
since it corresponds to unconstrained chains.

\begin{figure}[htbp]
\begin{center}
\includegraphics[scale=0.6]{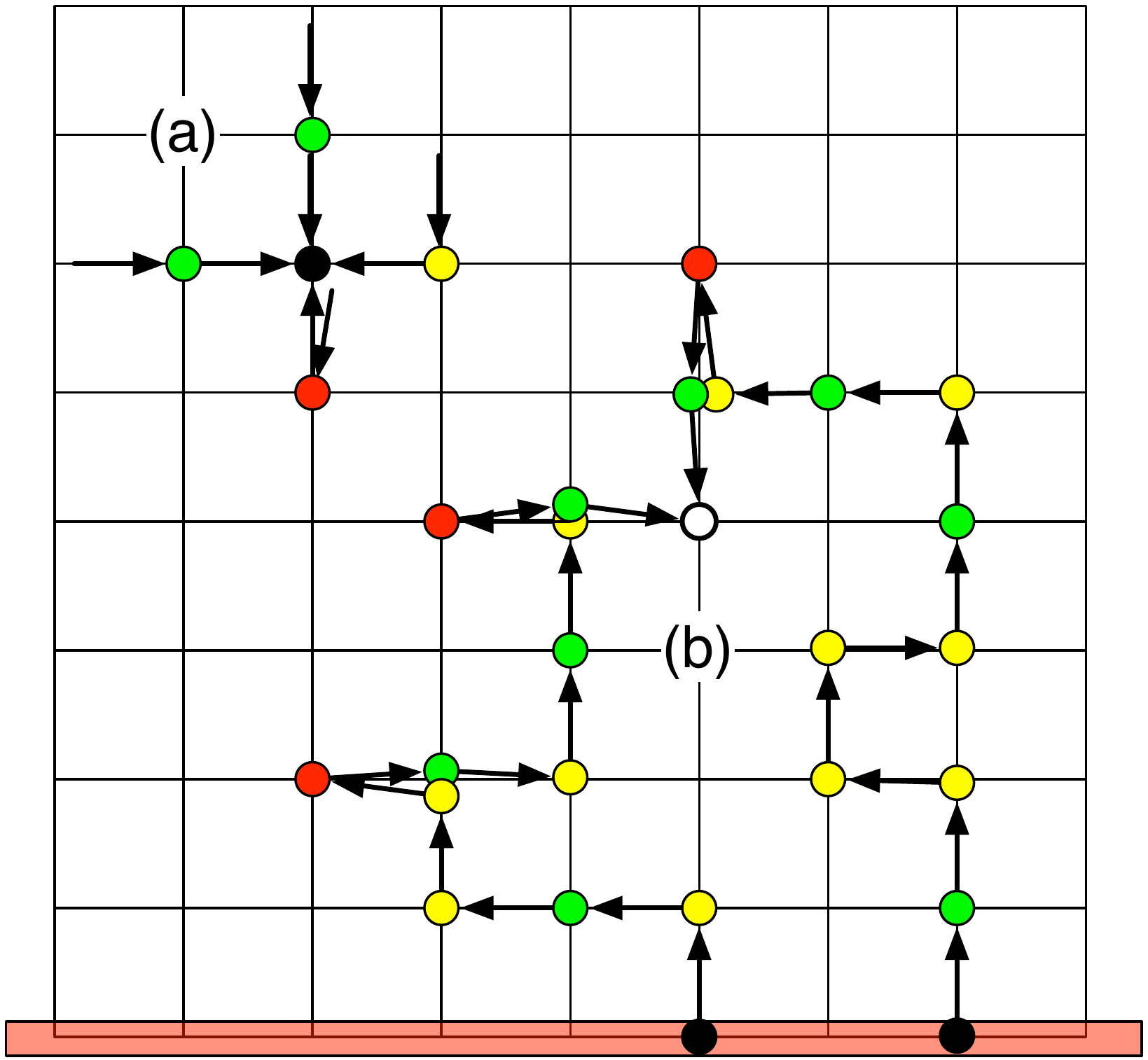}
\caption{a) Cartoon of lattice SCFT evolution; for a chain path to arrive at the central point (dark circle) on the $j$th step, it must have come from a neighboring site on the previous step.  b) The density at a given point (open circle) is the Boltzmann-weighted sum over all paths passing through the point.  Colors denote straight paths (green), bends (yellow), backfolds (red).}
\label{SCFT}
\end{center}
\end{figure}

SCFT consists of the following elements (see Fig.\ \ref{SCFT}):
\begin{enumerate}
\item make an initial guess for the self-consistent pressure potential 
$U_k$ on the $k$th layer;
\item given $U_k$, compute the Boltzmann weight $P(k,j)$
(single-chain partition function) for a chain segment
starting from the crystal surface to reach the $k$th lattice layer in $j$ steps;
\item compute the monomer density $\rho_k$ on the $k$th layer
(by summing over all paths through that layer, each weighted by its Boltzmann weight);
\item adjust $U_k$ until the density constraint (constant $\rho_k$ on each layer) is satisfied. 
%% changed
\end{enumerate}
(See Appendix for details.)

%% First portion of "methods" section returns here

Surprisingly, only the potential in the first layer is appreciably nonzero ---
$U_k$ for $k>1$ are smaller by a factor of typically 100 or more, and oscillatory functions of $k$.
This result suggests {\it assuming} that a nonzero value of $U_1$
is all that is required to achieve a self-consistent solution everywhere above the lamella.
Then, an arbitrary chain trajectory in the self-consistent potential
would consist of a sequence of ``trains'' (random walks in the first layer only)
and ``free loops'' into layers 2 and above.
(See Fig.\ \ref{one_layer_results}a.)
Because the potential $U_{k>1}$ is assumed zero,
the free loops would indeed be isotropic random walks above an absorbing boundary;
i.e., the same configurations of loops as in the isotropic melt pseudobrush.

Upon reflection, it is evident that one more constraint parameter is required;
namely, a Lagrange multiplier to control the injection density into layers 2 and above.
The single parameter $U_1$ cannot control both the melt density in layer 1,
and also the injection density into the space above.
One may say that this multiplier takes care of what those small oscillatory values 
of $U_{k>1}$ were doing in the numerical SCFT calculation.
With this addition, it turns out to be possible to enumerate chain configurations exactly;
the Boltzmann weight for a train is a geometric series. 
Likewise, a sequence of trains and loops results in a geometric series.

After considerable algebra, a startlingly simple formula emerges
for the free energy per loop $F(\sigma)$ as a function of loops per area $\sigma$:
%equation for F/loop, U1
\begin{eqnarray}
F/L & = & \left(\frac{5}{6 \sigma} - 1 \right) \log \left( \frac{5}{4} - \frac{3 \sigma}{2} \right) 
+ \log \sigma \nonumber \\
U_1 & = & -\log \left[ (5/4)(1-6\sigma/5) \right]
\label{eq4}
\end{eqnarray}

The free energy per loop (see Fig.\ \ref{one_layer_results}, lowest dark curve)
shows the expected minimum at the coverage $\sigma=1/6$,
which corresponds to the isotropic melt pseudobrush.
For $\sigma>1/6$, the potential $U_1$ tends to drive monomers out of the first layer,
while for $\sigma<1/6$ it tends to draw them in.
Note that the maximum possible coverage for flexible chains is $\sigma=5/6$;
at this coverage, on average for every six surface sites,
four injected chains have the tightest possible loop (one backfold)
and one makes a free loop in the space above the first layer.

\subsection*{Bending stiffness}
Of course, real chains have bending stiffness; 
on the lattice, 90 degree bends and backfolds (180 degree bends) should come at a price.
Fortunately, this can be incorporated into the lattice SCFT in a straightforward manner.
For semiflexible lattice chains, 
it suffices to compute the Boltzmann weight $P(k,t,j)$
for a chain to arrive at layer $k$ pointed in direction $t$ (either up, down, or horizontal)
after $j$ steps from the grafting surface.

%% second "Methods" section

Again, we can numerically solve the lattice SCFT equations for semiflexible chains.
(See Appendix for details.)
And again, we find the surprising result that the self-consistent potential
is essentially only one layer deep.
And so we are again motivated to assume this is so 
(and as before, we use a Lagrange multiplier 
to control the injection density of free loops into layer 2 and above).

Amazingly, even for semiflexible chains it turns out to be possible
to enumerate the relevant Boltzmann weights exactly,
by summing over chain configurations as sequences of trains and free loops.
(However, the resulting equation for the constraint potential $U_1$ 
must be solved numerically, so analytical expressions for the free energy per loop
cannot be given.)

\begin{figure}[htbp]
\begin{center}
\includegraphics[scale=0.6]{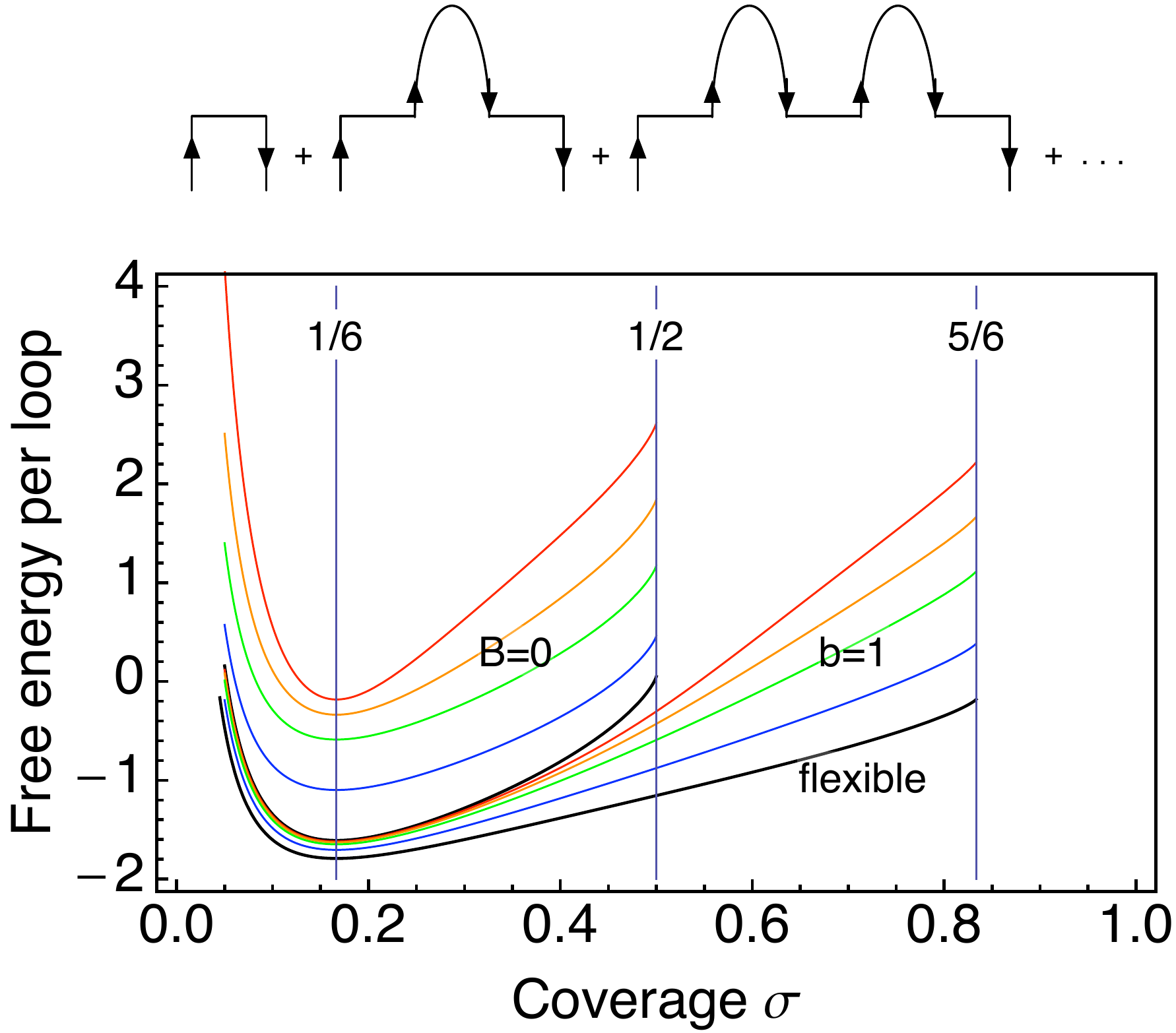}
\caption{a) General trajectory for a loop as a sequence of ``trains'' in the first layer, separated by ``free loops'' in the layers above (where $U=0$).  b) Free energy per loop, for various values of $b$ and $B$.  The lowest curve is the flexible limit, Eqn.\ (\ref{eq4}).  Immediately above, $b=1$ (no cost for 90 degree bend) and $B=0.5, 0.2, 0.1, 0.05$ (progressively stiffer backfolds).  Middle dark curve is $b=1, B=0$ (no backfolds); above are $B=0$ and $b=0.5, 0.2, 0.1, 0.05$ (progressively stiffer 90 degree bends).}
\label{one_layer_results}
\end{center} 
\end{figure}

%% added, explanation of parameters B and b
The semiflexible model has two parameters describing the bending penalties;
namely, the Boltzmann weights $b=\exp(-\beta \epsilon_b)$ for a 90 degree bend, 
and $B= \exp(-\beta \epsilon_B)$ for a 180 degree backfold.
%% end of addition
The resulting free energy per loop, for various values of $b$ and $B$, 
is displayed in Fig.\ \ref{one_layer_results}.
The upper set of curves correspond to $B=0$ (no backfolds allowed)
and increasing values of $b$ from top to bottom, 
with $b=1$ (no cost for 90 degree bend) as the last curve in the set.
The lower set of curves correspond to $b=1$ 
and increasing values of $B$ from top to bottom,
with $B=1$ (flexible limit) as the last curve.

Note that if $B$ is strictly zero, the maximum injection density is $\sigma=1/2$,
for which we have on average on every six surface sites,
two chains injected making a tight loop (two 90 degree bends and out),
and one chain making a free loop in the space above.

\subsection*{Putting polymers on a lattice}
To make contact with experiment, we map real polyethylene chains onto a lattice, as follows.
We choose a monomer to be two CH${}_2$ groups, 
and the cell volume to be the volume of this monomer in a real melt at 145C (about 59.5 \AA${}^3$).
One gauche bond in a chain gives rise to a 60 degree bend,
and two gauche bonds in succession gives a 90 degree bend,
so our monomer is able to make a 90 degree bend.
%% added, reason for small lattice
(The reason for taking a somewhat finer lattice than usual, 
corresponding to two monomer units rather than an entire Kuhn length, 
is so that the important effects of bending stiffness and finite extensibility 
can be incorporated into the lattice description.)
%% end of addition
The resulting cell linear dimension is then 3.9\AA.
%and the plaquette area is 14.47\AA${}^2$.

We forbid backfolds for steric reasons, and so set $B=0$.
We choose $b$ so that the mean-square end-to-end distance of chains on the lattice, 
given by 
\begin{equation}
\frac{R^2}{N} = a^2 \left( \frac{1+4b+B}{4b+2B} \right)
\label{eq8}
\end{equation}
matches the value for real melt polyethylene 
(about 1.25 \AA${}^2$/(g/mol))
\cite{FETTERS:1994p1446}.
This gives $b=0.192$ or $\epsilon_b = 5.7$kJ/mol at 145C;
this value is a bit lower than twice the trans-gauche energy difference (of about 3 kJ/mol),
but one can argue that short-distance fluctuations renormalize the naive value downwards.

The area per stem in the crystal phase is about 18.5\AA${}^2$,
which is a bit larger than the plaquette size in our lattice model.
However, we insist that the areal density of injected chains from the crystal
(half the areal density of chain stems) 
must correspond to the maximum coverage $\sigma=1/2$.
The argument for this identification is as follows:
if we injected chains at exactly twice this density, there would be no freedom whatsoever --- 
the chains would be obliged to be completely straight,
which would correspond to a continuation of the crystal.

% rearranged;  all amorphous-phase contribs discussed together
%
\subsection*{Amorphous-phase tension; interfacial tilt}
Real crystal-melt interfaces in polyethylene are commonly observed to be tilted 
with respect to the plane normal to the chain stems.
The most frequently observed interface is around [201], at about 34 degrees off normal
\cite{WUNDERLICH:1976}.
Intuitively, the interface tilts to relieve crowding of the chains entering the melt region;
but what limits the degree of tilt?

%% Third "Methods" section

To take advantage of the reduced crowding afforded by a tilted interface,
the chains must bend as they cross the interface
(so that they enter the melt region in the normal direction).
We model the bending cost as quadratic in the angle,
with a magnitude determined by the observation that at an inclination of 60 degrees,
every stem crossing the surface would require one gauche bond.

With this prescription, we can minimize the free energy per loop
with respect to the interfacial tilt angle,
and find the optimum tilt angle and corresponding interfacial free energy per area.
The results are in encouraging agreement with the common occurrence of [201],
also seen in atomistic simulations of PE
\cite{Gautam:2000p3340}.
We find an optimum tilt angle of 34 degrees, 
and a corresponding interfacial tension contribution (with respect to the base area)
of 22.2 mN/m, or about 1.5 kT per loop.
The fraction of adjacent reentry (defined for the lattice model as injected chains
that immediately make two 90 degree bends and reenter the lamella) 
is found to be 0.38, in good agreement with simulations
\cite{Balijepalli:2000p3338}.

Using our methods, 
the interface between a lamellar region of RII rotator phase and the adjacent melt
can be considered as easily as the crystal-melt interface.
The only difference between the two is a lower injection density of chains,
lower by a factor 18.25/21, the ratio of chains per area in the two ordered phases.
But because the crystal is so crowded, corresponding
to the maximum possible injection density ($\sigma = 1/2$),
the effect of a small increase in area per chain is significant.
We find that, in contrast to the crystal-melt interface,
the RII-melt interface occurs at essentially zero tilt angle, 
with an interfacial tension almost 40 percent lower --- 
14.7 mN/m versus 23.7 mN/m for the crystal.

We can also treat the interfacial free energy cost of the ``side'' 
of a crystal lamella adjacent to melt.
Indeed, this corresponds to the limit of zero injected chain density,
simply an impenetrable boundary from which isotropic chains are repelled.
In this case, the contributions is computed to be 10.1 mN/m and 8.9mN/m
for the crystal and rotator phases respectively.

%new section, on dispersive and melting contributions
\subsection*{Additional contributions to $\sigma$}
So far, we have focused on the interfacial free energy of the amorphous loops.
There are three additional contributions to the interfacial tension of the end and side faces, from
1) disruption of crystalline order by randomly reentrant loops on the end faces; 
2) missing ordered-phase neighbors on the side faces; and
3) dispersive interactions between phases of different density.
In all three cases, the contribution for rotator-melt interfaces is smaller than for crystal-melt interfaces,
because the the enthalpy and density differences with respect to the melt are smaller
for the rotator phase than for the crystal phase. 
(See Appendix for details).

The most important of these contributions 
is the interfacial free energy associated with disruption of crystalline periodicity,
at the interface in proximity to the amorphous region.
Atomistic Monte Carlo simulations of the Rutledge group\cite{Hutter:2006p3285} 
show that on the crystalline side of the interface 
(as defined by the density variation) the crystalline order is disrupted.

The potential energy rises sharply with a tanh-like profile,
with a width of about 1.6\AA\, the height of a monomer or so in the all-trans configuration
(see Figs.\ 3 and 4 of Ref.\ \cite{Hutter:2006p3285}).
We model this interfacial free energy arising from disordering the crystal using Cahn-Hilliard theory,
with parameters from Ref.\ \cite{Hutter:2006p3285} and experiment 
(see Appendix for details).

\begin{table}[htbp]
\begin{tabular}{c||r@{.}lr@{.}lr@{.}lr@{.}l|r@{.}lr@{.}lr@{.}lr@{.}l|r@{.}l||r@{.}lr@{.}l}
 & \multicolumn{8}{c|}{end} & \multicolumn{8}{c|}{side} & 
\multicolumn{2}{c||}{spherical} & \multicolumn{4}{c}{expt.} \\ \hline
phase  
& \multicolumn{2}{c}{xtal} & \multicolumn{2}{c}{melt} 
& \multicolumn{2}{c}{vdW} & \multicolumn{2}{c|}{Total} 
& \multicolumn{2}{c}{xtal} & \multicolumn{2}{c}{melt} 
& \multicolumn{2}{c}{vdW} & \multicolumn{2}{c|}{Total}
& \multicolumn{2}{c||}{$(\sigma_e \sigma_s^2)^{1/3}$} 
& \multicolumn{2}{c}{$(\sigma_e \sigma_s^2)^{1/3}$} 
& \multicolumn{2}{c}{$\sigma_e$} \\ \hline \hline
R &  18&3 & 14&7 & 0&2 & 33&1  & 7&3 & 8&9 & 0&2 & 16&4  & {\bf 20}&{\bf 7} 
& {\bf 23}&{\bf 5} & \multicolumn{2}{c}{---} \\
X &  32&0 & 23&7 & 1&4 & {\bf 57}&{\bf 1} & 13&0 & 10&1 & 1&4 & 24&5 & 32&5 
& 37&7 & {\bf 54}&{\bf 6} \\
\end{tabular}
\caption{Calculated contributions to end and side interfacial tension $\sigma_e$ and $\sigma_s$ 
for rotator (R) and crystal (X) phases, and ``spherical average'' $(\sigma_e \sigma_s^2)^{1/3}$; experimental values for spherical average and $\sigma_e$, described in text.}
\end{table}

Table 1 reports the computed values for all contributions to the melt interfacial tensions 
of rotator (R) and crystal (X) phases, on end and side faces:
``xtal'' are from disruption of order (end) or missing neighbors (side);
``melt'' are from free energy of amorphous loops (end or side, treated in the previous sections);
``vdW'' are from dispersive interactions.

There are no direct experimental values for any of the interfacial tensions,
except the end-face crystal-melt tension $\sigma_{XL}^{(e)}$,
obtained from the slope of the Gibbs-Thompson melting line
\cite{Strobl:2000p2183}.
Remarkably, the measured value is within 5\% of the computed result (see Table 1).
This agreement affords a small measure of confidence in our other interfacial tension values.

%completely new
\section*{Nucleation barriers}
With values for the various interfacial tensions in hand,
we set about estimating the barriers to nucleation via rotator and crystal phase nuclei.
For this purpose, we require values for the transition temperatures $T_{XL}$ and $T_{RL}$
and transition entropies $\Delta S_{XL}$ and $\Delta S_{RL}$ (X = crystal, R = rotator RII, L = liquid).

The transition entropies $\Delta S_{XL}$ and $\Delta S_{RL}$ 
are available from calorimetry experiments on linear alkanes in the range C${}_{20}$--C${}_{60}$
\cite{Kraack:2000p2178}.
The values of $\Delta S_{XL}$ and $\Delta S_{RL}$ depend on carbon number $n$, 
but settle down  by $n=35$ or so to essentially constant values of 0.718J/(g K) and 0.485J/(g K) respectively.
We obtain $T_{XL}=145$C by  extrapolating the ``melting line'' 
($T_{XL}(d)$ versus lamellar thickness $1/d$)
\cite{Strobl:2000p2183}.

$T_{RL}$ is more problematic,
since rotator phases are only metastable in polyethylene at standard pressure. 
Values from the alkane series have significant $n$-dependence even at $n=60$,
and extrapolation to $n=\infty$ is not unambiguous.
Instead, we make use of the value for $T_{XR}$ of 158C,
from extrapolating the crystallization line to infinite lamellar thickness
\cite{Strobl:2000p2183}.
Then, because the transition temperatures $T_{RL} < T_{XL} < T_{XR}$ are all close,
we determine $T_{RL}$ from a linear approximation on the free energy difference $\Delta F_{XR}$,
\begin{equation}
\Delta F_{XR} = \Delta F_{XL} - \Delta F_{RL}
  = (T - T_{XL}) \Delta S_{XL} - (T - T_{RL}) \Delta S_{RL}
\end{equation}
and the requirement that $\Delta F_{XR}=0$ at $T=T_{XR}$, 
which fixes the value of $T_{RL}=139$C.

\begin{figure}[htbp]
\begin{center}
\includegraphics[scale=0.8]{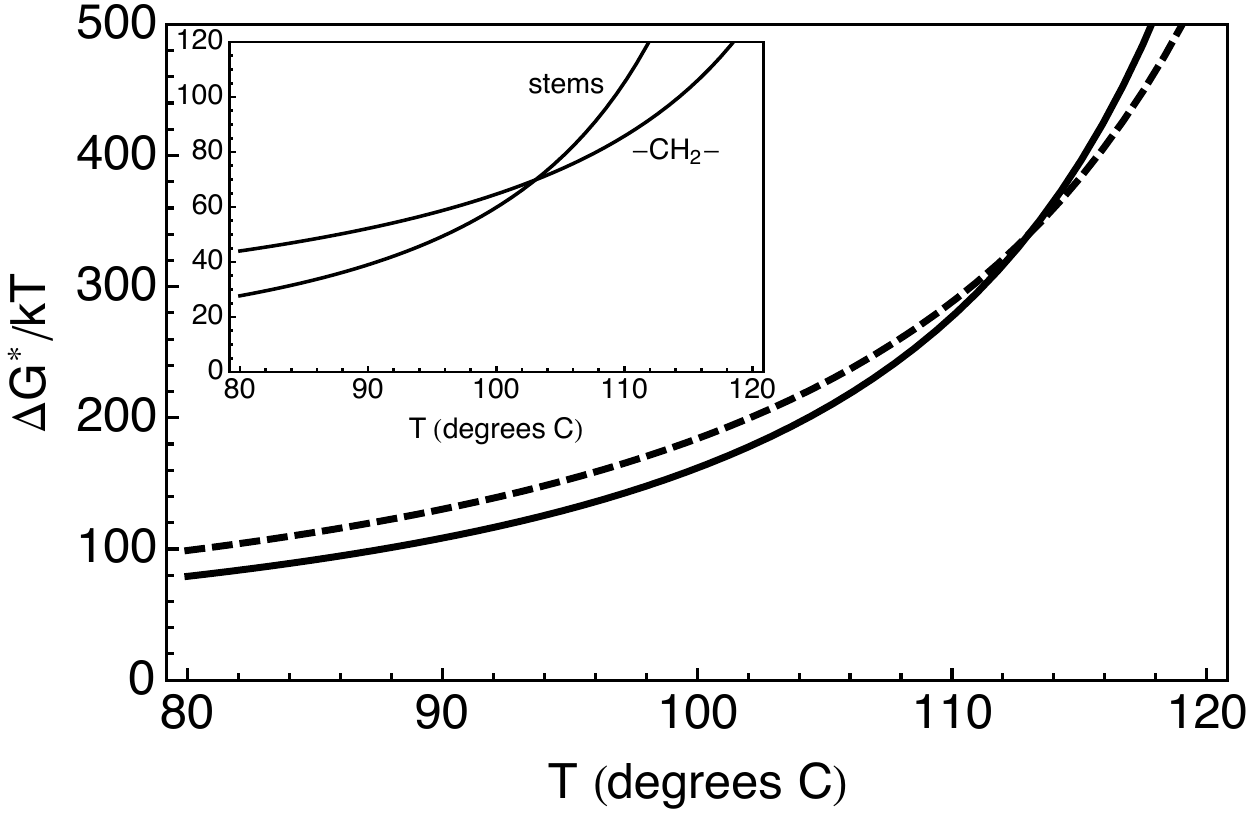}
\caption{\label{barrierfig}
Barriers for nucleation via rotator (solid) and crystal nuclei. 
Inset:  number of stems and CH${}_2$ groups.}
\end{center}
\end{figure}

Now, using Eqn.\ (\ref{nucl_barrier}) we can compute $\Delta G^*$ as a function of temperature,
assuming nucleation via rotator or crystal phase.
The result (see Fig.\ \ref{barrierfig}) is satisfying:
the barrier is lower for nucleation via rotator, for temperatures lower than about 115C.
Nucleation theory also predicts the area and thickness of the critical nucleus
\cite{Kraack:2000p2178},
shown in the inset to Fig.\ \ref{barrierfig}.
Above 115C, both barriers increase sharply, and indeed nucleation is essentially unobservable 
at such shallow undercoolings in PE.
The rotator phase wins out both because its interfacial tensions are lower,
and because it is nearly stable, with $T_{RL}$ only 6C below $T_{XL}$.

Our nucleation theory can also be compared to results of Ross and Frolen,
who measured homogenous nucleation rates in PE
\cite{ROSS:1975p4200}.
Their data can be compared to the classical prediction for the nucleation rate,
\begin{equation}
\log I(T) = \log K_\nu - D/(T \Delta T^2)
\end{equation}
in which the slope $D$ on a plot of $\log I(T)$ versus $1/(T \Delta T^2)$ is predicted to be
\begin{equation}
D = \frac{8 \pi \sigma_{e} \sigma_{s}^2}{k \Delta S^2}
\end{equation}

Thus, a value for $(\sigma_{e} \sigma_{s}^2)^{1/3}$ can be extracted from the Ross-Frolen data,
if one assumes the critical nucleus to be either crystal or rotator phase
(therefore determining the appropriate values of $\Delta S$ and $\Delta T$).
Table 1 reports best-fit values for $(\sigma_{e} \sigma_{s}^2)^{1/3}$ under both assumptions,
in which the log attempt frequency $\log K_\nu$ was also fitted.
The Ross-Frolen data itself is not sufficient to distinguish which phase nucleates,
but it is noteworthy that the inferred interfacial tension values for nucleation via rotator phase
are within 13\% of our predictions.
Finally, the fitted attempt frequency, 
long estimated to be the vibrational time of a Kuhn segment (about $10^{34}$/sec)
\cite{ROSS:1975p4200},
is ``only'' $10^3$ times this nominal value assuming rotator phase nucleation, 
but is fitted to be nearly $10^8$ times larger if we assume nucleation via crystal.

%% added comment on other rotator phases
Rotator phases may be operative in the crystallization of other polymers as well.
The characteristic behavior of the melting and crystallization lines seen by Strobl for PE
is repeated in data for isotactic and syndiotactic polypropylene
\cite{Strobl:2000p2183,Iijima:2000p8107}.
Ordered packings of parallel stems (all-trans, or helical) are a common feature of polymer crystals,
and these stems could rotationally disorder in a fashion similar to PE.
Rotator phases are only well known in PE because of studies on alkanes;
oligomers of other polymers can also be studied experimentally or simulated
to look for metastable rotator phases, a subject of continuing work in our group.
It may be that some aspects of the present approach
could also be relevant to crystallization of colloidal particles or proteins,
where densely packed but rotationally disordered mesophases 
may also play an important role in nucleation.
%% end of addition

{\it Acknowledgements:} I thank Eric Sirota, Gert Strobl, and Elie Raphael for useful discussions, 
and the Laboratoire ``GULLIVER'' at the Ecole Superieure de Physique et de Chimie Industrielles 
de la Ville de Paris (ESPCI) for their hospitality during which a portion of this work was carried out.  
This work was supported in part by NSF DMR-0907370.

\section*{Appendix}
\subsection*{Self-consistent field theory, flexible case.}
The equations that implement this procedure are 
\begin{eqnarray}
P(k,n+1) &=& (1/6) \left[ P(k-1,n) + 4 P(k,n) + P(k+1,n) \right] \exp(-U_k) \nonumber \\
P(k,1) &=& \delta_{k,1}
\label{eq1}
\end{eqnarray}
\begin{eqnarray}
\delta \rho(k,n) & = & \exp(U_k) \sum_j P(k,j) P(k,n-j) \nonumber \\
\rho(k) & = & \sigma \sum_n \frac{ n p(n) \delta \rho(k,n)}{\sum_k \delta \rho(k,n)}
\label{eq2}
\end{eqnarray}
The normalization of the Boltzmann weights so calculated
is such that free chains have a Boltzmann weight of unity.

Here we have introduced loop length distribution $p(n)$,
the probability that an injected chain will form a loop of length $n$.
The loop length distribution for the crystal-melt interface
is determined to minimize the free energy per loop $F$, given by
%F/loop, as sum of three terms
\begin{equation}
F/L = - \sum_n p(n) \log P(1,n) +  \sum_n p(n) \log p(n) - (1/\sigma) \sum_k U_k
\label{eq3}
\end{equation}

The first term is the $p(n)$-weighted average of the free energy 
(log of the partition function) for a loop of length $n$;
the second is the Shannon entropy of the loop length distribution itself.
[The third term subtracts the (artificial) work done by the constraint potential.]
Minimizing over $p(n)$ gives the reasonable result 
that $p(n)$ is proportional to the Boltzmann weight for loops of length $n$.

The resulting numerical procedure can be carried out for a melt-crystal interface
at different values for the ``injection density'' 
(areal density of chains emerging from the lamella).
(Practically, the self-consistent potential $U_k$ is computed stepwise,
by enforcing the density constraint on only the first layer using only $U_1$,
then on the first two layers using only $U_1$ and $U_2$, and so forth,
each time using the previous result for $U_k$ as the initial guess.)

\subsection*{Semiflexible case}
In terms of the Boltzmann weight $P(k,t,j)$, the generalized SCFT equations can be written
\begin{eqnarray}
P(k,+,n+1) & = & C \left[ P(k-1,+,n) + b P(k-1,0,n) + B P(k-1,-,n) \right] 
\exp(-U_k) \nonumber \\
P(k,0,n+1) & = & C \left[ (1+2b+B)P(k,0,n) + 4b P(k,+,n) + 4b P(k,-,n) \right] 
\exp(-U_k) \nonumber \\
P(k,-,n+1) & = & C \left[ P(k+1,-,n) + b P(k+1,0,n) + B P(k+1,+,n) \right]
\exp(-U_k) \nonumber \\
C & = & 1/(1+4b+B)
\label{eq5}
\end{eqnarray}
with corresponding expression for the density 
\begin{equation}
\rho(k, n) = \exp(U_k) \sum_{m=1}^n \sum_{t_1,t_2} P(k,t_1,m) P(k,t_2,n-m) M_{t_1 t_2}
\label{eq6}
\end{equation}
with $t_1$ and $t_2$ taking on the values $+, -, 0$, and the coupling matrix $M$ given by
\begin{equation}
M = \left[ 
\begin{array}{ccc}
B & b & 1 \\
b & (1+2b+B)/4 & b \\
1 & b & B 
\end{array}
\right]
\label{eq7}
\end{equation}
In the above, $b = \exp(-\beta \epsilon_b)$ and $B = \exp(-\beta \epsilon_B)$
are respectively the Boltzmann factors for a 90 degree bend and a backfold.
(See Fig.\ \ref{SCFT}.)

% new section
\subsection*{``Disordering'' contribution to interfacial tension} 
We model this interfacial free energy arising from disordering the crystal using Cahn-Hilliard theory.
We write an effective free energy per unit interfacial area as
\begin{equation}
\beta F/A = (1/\nu) \int dz \, \left( \frac{a^2}{2} \left( \frac{\partial \psi}{\partial z} \right)^2 
+ 16b \psi^2(1-\psi)^2 \right)
\end{equation}
Here $\nu$ is a monomeric volume, $a$ is a characteristic length, 
and $\psi$ is the order parameter (e.g., the strength of the density wave).
The second term is the usual quartic double-well potential,
with minima at $\psi=1$ (ordered state) and at $\psi=0$ (disordered state).
At the equlibrium melting point, the two states have equal free energy.
The height of the double-well barrier is $b$, which is the free energy cost at $\psi=1/2$, 
``halfway between'' the ordered crystal and disordered state.

The Cahn-Hilliard solution gives the order parameter profile $\psi(z)$ as a tanh,
\begin{equation}
\psi(z) = (1/2)\left( 1 + \tanh(4\sqrt{2b}z/a) \right)
\end{equation}
with a width $w=a/(4\sqrt{2b})$, which we set equal to 1.6\AA.
The interfacial free energy is 
\begin{equation}
\beta F/A = 16bw/(3 \nu)
\end{equation}

The free energy density at the interface midpoint is $b/(4 \nu)$,
so $\beta F/A$ equals $(4/3) f_{max} w$,
scaling as the maximum free energy density penalty times the width.
We make a guess for the ``barrier height'' (free energy value at $\psi=1/2$),
that it is half as big as the maximum possible value, i.e., half the transition enthalpy.
(The maximum barrier would obtain if the intermediate state 
had none of the energy reduction of the ordered phase,
and none of the entropy gain of the ordered phase.)

With values for transition enthalpies and temperatures as described in the main text.
and densities of 0.99g/cm${}^3$ and 0.91g/cm${}^3$ for crystal and rotator phases,
the resulting energy densities are 300J/cm${}^3$ and 171J/cm${}^3$ respectively.
We assume the interfacial widths to be the same for the crystal and rotator interfaces,
namely $w=1.6$\AA, and take $\nu$ to be the volume of a monomer in the ordered phase.
Applying these results gives the disordering contributions to $\sigma_{end}$ of Table 1.

%"broken bond" estimate, for side tension
On the sides of the nucleus, crystalline order terminates abruptly,
without the disordering effect of returning chains.
Here, a sensible estimate of the short-range contributions to interfacial energy
is to say that bonds to missing ordered-phase neighbors on the surface are ``broken''.
The interfacial energy cost is the formation enthalpy per CH${}_2$ group,
divided by the surface area of a cell containing one CH${}_2$ group,
times the fraction of cell area associated with missing neighbors.
This procedure gives contributions reported in Table 1.

%restored; was in supplementary
\subsection*{Dispersive contribution to interfacial tension}
The density of the melt (about 0.845g/cm${}^3$) at room temperature 
is slightly lower than that of the crystal (about 1.0g/cm${}^3$)
\cite{BRANDRUP:1999}.
Thus, the density of polarizable groups giving rise to dispersive forces, 
and hence the Hamaker constant, would be larger in the crystal than the melt.
The rotator phase has a density we estimate from the ratio of areal densities for crystal and rotator,
as 18.5/21 times 1g/cm${}^3$, or about 0.9g/cm${}^3$.

The Hamaker constant can be well estimated from the index of refraction 
using an approximation to the full Lifshitz theory of dispersive interactions
\cite{ISRAELACHVILI:1992}:
\begin{equation}
A \approx \frac{3 h \nu_e}{16 \sqrt{2}} \frac{(n_1^2 - n_2^2)^2}{(n_1^2 + n_2^2)^{3/2}}
\end{equation}
in which $\nu_e$ is a typical optical absorption frequency of about $3 \times 10^{15}$Hz,
$n_1$ and $n_2$ are the indices of refraction of the two phases.

The difference in index of refraction between crystal and melt 
can be obtained from the Clausius-Mossotti relation
\cite{JACKSON:1975},
\begin{equation}
\frac{n^2 - 1}{n^2 + 2} \propto \rho
\end{equation}
The dispersive contribution to interfacial tension is then given
by an integral of the dispersive forces between two slabs of material A (crystal)
interacting across a slab of material B (melt), comparing results at large separations 
and at a short-distance cutoff of about 1.5\AA
\cite{ISRAELACHVILI:1992}:
\begin{equation}
2 \gamma = \frac{A}{12 \pi D_0^2}
\end{equation}

We find for liquid normal alkanes interacting across air, a value of
$A_{232} = 6.59 \times 10^{-20}$J (1 = ordered, 2 = liquid, 3 = air),
and for solid normal alkanes interacting across liquid normal alkanes,
a value of $A_{121} = 2.80 \times 10^{-21}$J.
Assuming the cutoff distance is the same in the two cases,
the interfacial tension $\gamma_{12}$ is then
\begin{equation}
\gamma_{12} = \gamma_{23} \frac{A_{121}}{A_{232}}
\end{equation}
Alkane interfacial tensions against air at 20C,
extrapolated to high molecular weight, gives $\gamma_{23} = 33.5$dyn/cm.
Applying these results gives the values labeled ``vdW'' in Table 1.

%%%%%%%%%%%%%%%%%%%%%%%%%%%%%%%%%%%%%%%%%%%%%%%%%%%%%%%%%%%%%%%%%%%%%
%% The appropriate \bibliographystyle and \bibliography commands
%% should be placed here.
%%%%%%%%%%%%%%%%%%%%%%%%%%%%%%%%%%%%%%%%%%%%%%%%%%%%%%%%%%%%%%%%%%%%%
\bibliographystyle{rsc}
\bibliography{melt-xtal}

\end{document}